\documentclass{llncs}
\usepackage{amsmath,amsfonts}
\usepackage{algorithmic}
\usepackage{algorithm}
\usepackage{array}
\usepackage[caption=false,font=normalsize,labelfont=sf,textfont=sf]{subfig}
\usepackage{textcomp}
\usepackage{stfloats}
\usepackage{url}
\usepackage{verbatim}
\usepackage{graphicx}
\usepackage{cite}
\usepackage{lineno,hyperref}
\usepackage{amsmath}
\usepackage{float}
\usepackage{amssymb}
\usepackage{amsfonts}
\usepackage{mathtools}
\usepackage{subcaption}
\usepackage{xcolor}
\usepackage[most]{tcolorbox}

\newtcolorbox{resultsummarybox}{
    enhanced,
    breakable,
    colback=gray!4,
    colframe=gray!55,
    colbacktitle=gray!12,
    boxrule=0.6pt,
    arc=2pt,
    left=6pt,
    right=6pt,
    top=6pt,
    bottom=6pt,
    title={Main takeaways},
    fonttitle=\bfseries,
    coltitle=black
}

\begin{document}

\title{Towards National Quantum Communication in Europe: Planning and Sizing Terrestrial QKD Networks}

\author{Sebastian Raubitzek\inst{1} \and Werner Strasser\inst{2} \and Sebastian Ramacher\inst{3} \and Thomas Lebeth\inst{4} \and Andreas Neuhold\inst{5} \and Christoph Pacher\inst{2,3,6}}
\institute{SBA Research gGmbH, Vienna, Austria \and
fragmentiX Storage Solutions GmbH, Vienna, Austria \and
AIT Austrian Institute of Technology GmbH, Vienna, Austria \and
dacoso GmbH, Vienna, Austria \and
CANCOM Converged Services GmbH, Vienna, Austria \and
Christian Doppler Laboratory AsTra, Vienna, Austria}

\maketitle

\begin{abstract}
The European Union is developing the European Quantum Communication Infrastructure (EuroQCI) as a pan-European network to provide secure communication capabilities across Member States, including governmental and critical-infrastructure domains. While the strategic objective is defined at EU level, the required scale and structure of national quantum key distribution (QKD) networks remain largely unspecified.

This work addresses the question of how to plan and size national terrestrial QKD networks to support critical infrastructure and public authorities. We propose a reproducible planning methodology that estimates network size, total fiber length, and the number of required QKD components based on a small set of explicit assumptions. The approach is demonstrated for Austria, where a synthetic but structured network model is constructed and evaluated using Monte Carlo simulation.

The model focuses on terrestrial QKD infrastructure and explicitly excludes space-based segments. It estimates endpoint counts, trusted repeater node requirements, and hop-length distributions under realistic operational constraints. The Austrian case is then used as a baseline to derive scaling rules for other EU Member States based on population and geographic extent.

The results provide first-order planning estimates for national QKD backbone sizes across Europe. These estimates are not intended as deployment designs but as planning-level references that support early-stage cost assessment and infrastructure dimensioning under the EuroQCI framework.
\end{abstract}


\section{Introduction}

The European Union is establishing the European Quantum Communication Infrastructure (EuroQCI) as a secure communication network spanning all Member States \cite{EC_EuroQCI}. The objective is to interconnect governmental institutions, critical infrastructures, and other security-relevant entities using quantum key distribution (QKD) technologies. EuroQCI is intended to operate as a continuous and maintained infrastructure, integrating national terrestrial networks with cross-border links and, in a later stage, satellite-based components \cite{CEF_EuroQCI_CallFiche_2024}.

At the policy level, the need for secure communication in critical infrastructure is reinforced by regulatory frameworks such as the NIS2 Directive and the CER Directive, which require Member States to ensure resilience and cybersecurity across essential services \cite{NIS2_2022_2555,CER_2022_2557}. These requirements imply the need for secure key distribution mechanisms across a wide range of geographically distributed sites, including public administration sites, governmental institutions, and other critical entities.

Despite these strategic objectives, a fundamental planning question remains open: \emph{how large must national QKD networks be to provide sufficient coverage and connectivity?} In particular, it is unclear how many nodes are required per Member State, how densely these nodes must be interconnected, and what total infrastructure length is needed to support a resilient network. Existing work on QKD networks focuses primarily on technological feasibility, architectural concepts, and specific deployments \cite{Stanley2022_QKDNetworks,ETSI_QKD_002,ETSI_QS_Whitepaper,DBLP:conf/ofc/LopezBPMSRL21}, and only limited guidance is available for generalizable nationwide infrastructure sizing.

This work addresses this gap by proposing a planning-level methodology to estimate the size of terrestrial QKD networks at national scale. The approach is based on a synthetic network model that represents endpoints corresponding to security-relevant sites and constructs a moderately meshed backbone under explicit constraints. The model incorporates practical QKD limitations, in particular the need to limit hop lengths and the use of trusted repeater nodes to extend communication distances \cite{Huttner2022}.

The methodology is demonstrated using Austria as a reference case. Based on expert-informed assumptions, a representative endpoint distribution is generated and a corresponding backbone network is constructed. Monte Carlo simulation is used to account for variability in spatial placement and network realization. From this, estimates for total fiber length, hop-length distributions, and required component counts are derived. The Austrian case is then used to define rough scaling rules that allow transfer of the approach to other Member States.

It is important to note that the presented results are \emph{planning estimates} rather than deployment specifications. Even for a single country, the exact network realization depends on detailed factors such as existing fiber infrastructure, sector-specific requirements, and operational constraints. The proposed approach therefore aims to provide a transparent and reproducible first estimate based on clearly stated assumptions.

Space-based QKD extends secure key distribution beyond the distance range achievable in terrestrial fibre networks. 
By providing QKD to geographically isolated regions—such as the Azores, French Guiana, and Reunion—and overseas embassies, space-based QKD ensures total territorial sovereignty, but requires specialized infrastructure such as satellites and optical ground stations and is subject to additional operational constraints \cite{Bedington2017_SatQKD,Sidhu2023_SatQKD}.
In this context, space-based QKD is best treated as a complementary component of a hybrid architecture rather than a substitute for a terrestrial backbone. This is consistent with the EuroQCI approach, which combines terrestrial and space segments while accounting for the broader operational and security dependencies of satellite systems \cite{EC_EuroQCI,ENISA_Space_Threat_Landscape_2025,NISTIR8401}.
 
Due to the high capital expenditure of ground stations, limited link availability, and lower key-exchange bandwidth, the majority of QKD traffic will remain terrestrial. Consequently, this analysis restricts its scope to the EuroQCI terrestrial segment.

%The analysis focuses exclusively on the terrestrial segment of EuroQCI. While space-based QKD can provide long-distance links, satellite systems introduce additional dependencies on space infrastructure, including launch services, orbital assets, ground control segments, and telemetry and command chains. These systems are subject to cybersecurity and operational risks across their lifecycle, including ground-origin cyber attacks, spoofing and hijacking of communications, supply-chain compromise, and physical compromise of ground-based and pre-launch assets \cite{ENISA_Space_Threat_Landscape_2025,NISTIR8270,NISTIR8401}. In addition, satellites themselves are physically vulnerable to counterspace capabilities, including direct-ascent anti-satellite (ASAT) systems that have been demonstrated to destroy satellites in orbit, resulting in immediate loss of functionality and long-lived debris \cite{SecureWorld2022_ASAT,ESA_SpaceDebris_ASAT}.

From a network-planning perspective, terrestrial infrastructure 
%remains the primary focus because it 
enables continuous operation, direct control, and more accessible maintenance and recovery. 
%At the same time, purely terrestrial QKD introduces challenges for geographically complex Member States, in particular for countries with distant islands such as Spain or Portugal, where long-distance maritime links are difficult to realize with trusted repeater nodes.

Based on these considerations, this work formulates the following research questions:

\begin{itemize}
    \item How many QKD endpoints and trusted repeater nodes are required to support a national critical-infrastructure backbone?
    \item What total fiber length is implied by such a network under realistic geographic and operational constraints?
    \item How can these quantities be systematically estimated and scaled across EU Member States?
\end{itemize}

By addressing these questions, we provide a consistent framework for early-stage dimensioning of national QKD infrastructures within the EuroQCI context.

\begin{tcolorbox}[title=Problem Statement,colback=gray!10,colframe=gray!50]
\begin{itemize}
    \item Planning-level estimation of national terrestrial QKD networks within EuroQCI.
    \item Focus on reproducible sizing under realistic constraints, not deployment design.
    \item Research objectives:
    \begin{itemize}
        \item number of endpoints and trusted repeater nodes,
        \item total fiber length,
        \item scaling across Member States.
    \end{itemize}
\end{itemize}
\end{tcolorbox}

The remainder of this article is structured as follows. Section~\ref{sec:related} reviews existing work on QKD networks, including trusted-node architectures, network design approaches, and the role of key management systems in large-scale deployments. Section~\ref{sec:austria_reference_case} defines Austria as the reference planning case and specifies the assumptions used to model a national QKD backbone. Section~\ref{sec:simulation_overview_assumptions} describes the simulation framework, including endpoint generation, graph construction, and trusted repeater placement. Section~\ref{sec:discussion_austria} presents the results for the Austrian case and analyzes the resulting network characteristics. Section~\ref{sec:scaling_from_austria_eu} extends the approach to other EU Member States and defines the corresponding scaling assumptions, while Section~\ref{sec:eu_results} reports the resulting cross-country simulation outcomes. Finally, Section~\ref{sec:discussion_conclusion} discusses the results, outlines the limitations of the approach, and summarizes the main conclusions and directions for future work.

%===============================================================
\section{Related Work}
\label{sec:related}
%===============================================================

Quantum key distribution (QKD) has evolved from point-to-point experimental systems to networked infrastructures intended for metropolitan and national deployment. Foundational protocol work established the feasibility of secure key exchange over optical fiber and free-space channels \cite{Bennett1984}. Subsequent experimental and field deployments demonstrated practical implementations and extended these systems into network configurations, including metropolitan testbeds and trusted-node architectures \cite{Peev2009,Dynes2019}. These developments highlight that practical QKD deployment requires integration into existing communication infrastructure and must address distance limitations, key management, and operational stability. Comprehensive reviews describe the transition from laboratory systems to field-deployed networks and outline the engineering constraints that shape real-world QKD infrastructures \cite{Scarani2009,Pirandola2020,Diamanti2016}. Recent large-scale review work on continuous-variable quantum communication further consolidates these developments by covering modern protocols, implementation techniques, and system-level considerations for scalable quantum communication infrastructures \cite{Usenko2026_CV_QC_Review}.

\paragraph{Trusted repeater architectures and distance limitations} A central architectural concept in current terrestrial QKD networks is the use of trusted repeater nodes to extend communication distances beyond the limits of direct optical transmission. In this model, intermediate nodes perform key relaying under a trust assumption, enabling multi-hop secure links across large geographic areas. This architectural approach is well established in deployed QKD networks and system-level studies \cite{Peev2009}. European initiatives, including SECOQC and subsequent national testbeds, have demonstrated the feasibility of such architectures in operational environments. Work on long-distance fiber-based QKD systems further highlights the physical transmission limits that motivate multi-hop designs \cite{Stucki2011}. These studies emphasize the importance of network topology, link budgeting, and system interoperability.

\paragraph{Network design and optimization approaches} Beyond technological feasibility, several studies address the design and optimization of QKD networks. These works typically formulate network planning as an optimization problem under constraints such as link loss, node placement, and cost. Existing approaches include graph-based formulations and cost-minimization strategies for QKD network deployment, often combined with classical network design techniques. However, many of these studies rely on detailed demand assumptions or specific deployment scenarios, which limits their applicability for early-stage national-scale planning. In addition, several widely cited works focus on large-scale demonstrations or specific architectures rather than generalizable planning methodologies, highlighting a gap between optimization-based designs and planning-level estimation.

\paragraph{Large-scale initiatives and backbone networks} Recent work has also considered large-scale and backbone-level QKD networks. National and continental initiatives such as EuroQCI and related projects aim to establish wide-area secure communication infrastructures \cite{EC_EuroQCI,CEF_EuroQCI_CallFiche_2024}. Research in this context focuses on scalable architectures, interoperability, and integration with classical communication systems \cite{Kozlowski2020}. Studies of backbone QKD networks highlight the role of geographic constraints, fiber availability, and resilience requirements, but they typically focus on specific deployment cases or technology validation rather than generalized sizing methodologies. In parallel, satellite-based QKD has been proposed to complement terrestrial networks for long-distance links, although it introduces different operational and security considerations \cite{Liao2017}.

\paragraph{Key management systems in QKD networks} Recent work has placed increasing emphasis on the role of key management systems (KMS) as a central component of scalable QKD networks. In contrast to early network demonstrations, where key handling was often treated implicitly, modern architectures explicitly separate quantum key generation from key management, distribution, and application integration. In particular, recent work has contributed to the design of KMS architectures for large-scale QKD networks, including functional decomposition, interface definitions, and integration with software-defined networking (SDN) environments \cite{James2023_KMS,ETSI_QKD_015,ETSI_QKD_018,DBLP:conf/icton/MartinBOBVSSACSSEDRPL23}. These approaches highlight that KMS functionality is required to aggregate keys from multiple QKD links, enforce key lifecycle policies, and provide standardized interfaces between QKD modules, network control layers, and applications.

\paragraph{Scaling, interoperability, and KMS challenges} More recent developments address the challenges of scaling QKD beyond point-to-point links toward network-wide deployment. Recent work proposes KMS designs that support inter-KMS communication, distributed key storage, and integration with post-quantum cryptographic mechanisms, thereby enabling hybrid security architectures \cite{James2023_KMS}. In parallel, European initiatives such as OPENQKD have demonstrated the need for interoperable KMS solutions in heterogeneous multi-vendor environments and real-world use cases \cite{OPENQKD2021}. These developments are complemented by recent surveys that identify key management as a primary bottleneck for large-scale QKD deployment, particularly in terms of resource allocation, interface standardization, and integration with classical network infrastructures \cite{Mehic2024_KMS_Survey}. Overall, current research establishes KMS as a core architectural layer that connects quantum key generation with operational network services and application-level security.

\paragraph{Multi-path security and trust distribution} Recent studies also extend the discussion from KMS architecture alone to the broader question of how larger trusted-node QKD networks can reduce architectural single points of failure. Valbusa \emph{et al.} provide a structured overview of multi-path approaches for QKD networks and show that network-level redundancy can be used to reduce the security impact of compromised trusted nodes by distributing trust over multiple paths rather than relying on a single forwarding chain \cite{Valbusa2025_MultipathQKD}. This is directly relevant for large terrestrial backbones because it links trusted-node placement and KMS coordination to end-to-end security assumptions at network scale.

\paragraph{Implications for large-scale QKD architectures} Taken together, these studies reinforce the view that large QKD networks are not determined only by optical reach and trusted repeater placement. They also depend on how keys are routed, synchronized, protected, and consumed across multiple administrative and technical layers. This strengthens the importance of KMS design in larger QKD infrastructures, especially when networks are expected to support multi-path communication, hybrid cryptographic operation, and heterogeneous deployment environments. It also shows that the transition from metropolitan demonstrations to larger operational networks requires closer integration between QKD links, control-plane functions, and application-layer key delivery. For the present article, this is particularly relevant because a planning methodology for national terrestrial QKD backbones should not only estimate endpoints and trusted nodes, but should also remain compatible with the architectural role of distributed KMS instances in large-scale and operationally diverse networks \cite{James2023_KMS,Valbusa2025_MultipathQKD,Bastos2025_DISCRETION_ICMCIS,Brito2025_DISCRETION_QCNC}.

\paragraph{Planning methodologies and modeling approaches}
In the broader context of communication-network planning, early-stage dimensioning is often carried out under incomplete or uncertain demand information using synthetic traffic assumptions and scenario-based planning methods. Recent work in optical networks, for example, applies multi-period planning under uncertain traffic development to estimate infrastructure requirements before detailed deployment data are available \cite{Hosseini2023_MultiPeriodPlanning}. These approaches are closely aligned with the present work because they support reproducible planning under incomplete information rather than exact deployment design. By contrast, although optimization and architectural studies for QKD networks are growing, planning-oriented methodologies for nationwide QKD backbone sizing appear to be less prominently represented in the literature considered here \cite{HernandezHernandez2025_qTDM_QKDN}.

\paragraph{Positioning of the present work} The present work builds on these strands of research by combining elements of QKD network architecture with planning-level modeling techniques. Unlike optimization-based designs that require detailed input data, the proposed approach uses a small set of explicit assumptions to generate synthetic but structured network instances. This enables reproducible estimation of network size and component counts across different countries while remaining consistent with known QKD constraints, such as limited hop length and the use of trusted nodes. The contribution is therefore positioned between technology-focused QKD studies and classical network planning approaches, providing a transparent framework for early-stage infrastructure dimensioning in the EuroQCI context.

\begin{tcolorbox}[title=Related Work,colback=gray!10,colframe=gray!50]
\begin{itemize}
    \item QKD research spans protocol foundations, deployments, and network architectures.
    \item Trusted-node networks are standard but driven by physical distance limits.
    \item Optimization-based designs exist but often rely on strong assumptions.
    \item KMS is a central component for scalable and operational QKD networks.
    \item Large-scale deployment requires integration across network, control, and application layers.
\end{itemize}
\end{tcolorbox}

\section{Austria as the reference planning case}
\label{sec:austria_reference_case}

Austria is used in this work as the reference planning case for a terrestrial QKD backbone for critical infrastructure and public authorities. Austria provides a well-defined national setting with a manageable geographic scale and a clear federal structure. At the same time, the policy context is aligned with the objectives of secure communication for public administration and critical infrastructure under EuroQCI, cybersecurity regulation, and resilience requirements \cite{EC_EuroQCI,CEF_EuroQCI_CallFiche_2024,NIS2_2022_2555,CER_2022_2557}. Austrian national strategy documents further emphasize the protection of critical infrastructure, the growing relevance of cyber threats, and the need for resilient and secure digital communication capabilities as part of a broader security approach \cite{AT_Security_Strategy_2024,AT_Cybersecurity_Report_2024}. In particular, the Austrian Security Strategy highlights digitalisation and critical infrastructure dependencies as central security factors, while the national cybersecurity report documents the growing threat landscape affecting public institutions and operators of essential services. This combination of strategic prioritization and observed threat dynamics makes Austria a suitable baseline for a planning-level QKD infrastructure study. The purpose of this use case is therefore not to define a deployable network, but to establish a transparent and reproducible national planning scenario.

The Austrian use case is interpreted as a national QKD backbone that provides key material for governmental and critical-infrastructure communication, including applications that may fall under classified or classifiable information handling up to EU SECRET \cite{EUCI_2015_444}. In practical terms, QKD links are assumed between security-relevant sites in order to generate symmetric keys, which are then distributed and managed through classical key management infrastructure. The model focuses on the physical and logical structure required to support such a service at national scale. It does not model traffic demand, specific applications, or service-level agreements. Instead, it provides a consistent framework to estimate required link lengths, the number of QKD systems, and supporting infrastructure.

The planning case assumes a fixed set of $N_{\mathrm{EP}}=250$ endpoint locations across Austria. Each endpoint represents a site where secure key material must be available, such as ministries, state authorities, public authorities, or other critical infrastructure sites \cite{EC_EuroQCI}. The endpoint set is synthetic and represents a demand proxy rather than a concrete list of facilities. Its spatial distribution follows expert-informed assumptions: a high concentration in Vienna, smaller clusters in the other federal state capitals, and an additional rural distribution to account for geographically dispersed critical assets. This reflects the scope of NIS2 and CER, where obligations apply across sectors and regions rather than only in major urban centers \cite{NIS2_2022_2555,CER_2022_2557}. In addition, two nodes are placed in St.~Johann im Pongau where a backup data center of the government is located. As a result, the Austrian use case represents a nationwide service requirement.

Connectivity between endpoints is realized through direct QKD links wherever distances allow. Each such link is implemented as a pair of QKD devices connected via optical fiber. For reliability and operational flexibility, endpoints are assumed to be connected to multiple other sites. In practical terms, most endpoints are assumed to host two or three QKD links to different neighboring sites. This reflects a basic redundancy assumption: endpoints should not rely on a single connection for key supply if alternatives can be provided. At the national level, this results in a moderately interconnected backbone structure that avoids isolated regions and reduces the likelihood that large areas depend on a single connection path.

A central constraint for terrestrial QKD connections is the limited transmission distance of QKD over optical fiber. The model assumes that practical QKD links operate within a range of approximately 40--100\,km per span, while the more specific objective to keep individual spans clearly below 75\,km is treated here as a planning choice intended to support stable operation and to limit optical loss \cite{Huttner2022}. For longer distances, trusted repeater nodes (TRNs) are introduced. These are intermediate secure sites that terminate one QKD link and initiate another, effectively splitting a long connection into multiple shorter spans. In the Austrian planning case, the number of trusted repeater nodes is fixed to $N_{\mathrm{TR}}=50$. These nodes are not demand endpoints; they serve only to enable long-distance connectivity and maintain feasible QKD operation. This reflects the practical requirement that nationwide QKD deployment relies on trusted-node architectures in the absence of quantum repeaters \cite{Huttner2022}.

Distances in the Austrian use case are computed using geodesic distances on WGS84 coordinates. These distances are interpreted as a lower bound for physical fiber lengths. To approximate real-world routing effects, such as cable paths following roads or existing infrastructure corridors, a constant detour factor of $\alpha=1.5$ is applied, resulting in route-corrected lengths $L_{\mathrm{route}}\approx \alpha L_{\mathrm{geo}}$. This provides a consistent approximation of deployment distances without requiring detailed knowledge of existing fiber infrastructure. The same correction is applied when evaluating QKD span lengths between endpoints and trusted nodes.

The Austrian use case also defines explicit assumptions for the required QKD and key management components. Each QKD link between two sites requires a complementary device pair. Trusted repeater nodes are assumed to host such device pairs for each connected span. Endpoints are assumed to host multiple QKD devices, corresponding to their multiple connections. In the planning case, 150 endpoints are assumed to host three QKD devices and 100 endpoints are assumed to host two QKD devices. This results in a total of 650 QKD devices at endpoints. Trusted repeater nodes contribute an additional 100 QKD devices, leading to a total of 750 QKD devices in the Austrian case.

For the management layer, the model assumes that each security-relevant site operates a key management system (KMS) instance. This includes both endpoints and trusted repeater nodes, resulting in $N_{\mathrm{KMS}}=300$. Each KMS is assumed to be supported by a hardware security module (HSM), leading to $N_{\mathrm{HSM}}=300$. In practical terms, every site participating in the QKD backbone is assumed to be capable of securely storing, processing, and distributing key material. These assumptions reflect a distributed key management architecture consistent with current QKD system designs \cite{ETSI_QKD_015}.

This assumption is consistent with implementations of the key management layer, where QKD networks have been integrated with centralized and distributed key management systems that interface with cryptographic applications and support controlled key handling through dedicated security components. In particular, the SECOQC framework and more recent work on large-scale QKD key management systems demonstrate the practical use of KMS architectures for key aggregation, distribution, and application integration in operational QKD networks \cite{Peev2009,James2023_KMS}.

Taken together, the Austrian use case defines a national QKD backbone in which endpoints are interconnected through multiple QKD links, long distances are bridged using trusted repeater nodes, and all sites are equipped with key management capabilities. The model provides a consistent and reproducible baseline for estimating network scale, link lengths, and component requirements. This baseline is then used in the following sections to derive comparable planning cases for other countries.

\paragraph{Summary of the Austrian planning case}
The Austrian reference case is defined by the following main assumptions:
\begin{itemize}
    \item a national QKD backbone serving public authorities and critical infrastructure, aligned with EuroQCI objectives;
    \item $N_{\mathrm{EP}}=250$ endpoint sites distributed across all federal states, with higher density in Vienna, smaller regional clusters, and a rural component;
    \item endpoints connected via multiple QKD links, typically two or three per site, to provide basic redundancy;
    \item a trusted-node architecture with $N_{\mathrm{TR}}=50$ trusted repeater nodes used to split long distances into feasible QKD spans;
    \item QKD span lengths assumed to be in the range of approximately 40--100\,km, with a planning target below 75\,km;
    \item geodesic distance calculation with a constant detour factor $\alpha=1.5$ to approximate real fiber routing;
    \item explicit component assumptions including 750 QKD devices in total, $N_{\mathrm{KMS}}=300$, and $N_{\mathrm{HSM}}=300$;
    \item a distributed key management architecture where each endpoint and trusted node hosts KMS and HSM functionality.
\end{itemize}

\begin{tcolorbox}[title=Austria Reference Case,colback=gray!10,colframe=gray!50]
\begin{itemize}
    \item $N_{\mathrm{EP}}=250$ endpoints across Austria.
    \item $N_{\mathrm{TR}}=50$ trusted repeater nodes.
    \item 2--3 links per endpoint for redundancy.
    \item Span length $\approx 40$--$100$\,km (target $<75$\,km).
    \item 750 QKD devices, 300 KMS/HSM instances.
\end{itemize}
\end{tcolorbox}

\section{Simulation and Assumptions}
\label{sec:simulation_overview_assumptions}

This section summarizes the simulation framework used to generate synthetic QKD network instances for Austria and the remaining European country cases. The implementation follows a fixed sequence of steps: (i) load a country boundary geometry, (ii) sample QKD endpoint locations within that geometry using a clustered-plus-rural mixture model, (iii) assign a prescribed degree sequence to the sampled endpoints, (iv) construct an undirected simple graph that satisfies this degree sequence while preferring short connections, (v) generate multiple valid candidate graphs and retain the best one under a robustness-oriented score, (vi) place trusted repeater nodes (TRNs) on selected edges under a fixed budget, and (vii) export edge tables, node tables, plots, and summary reports. The full implementation is available in the project repository \href{https://github.com/Raubkatz/QCI_Simulation}{github.com/Raubkatz/QCI\_Simulation}. The present description follows the implemented simulation logic at methodological level, without relying on code-specific details. The simulation workflow described in this section is based on the Python implementation used for the present study and the associated generated report files. 

\paragraph{Geospatial domain and geometry assumptions}
The spatial domain for each simulation is given by a Natural Earth country polygon or multipolygon in WGS84 coordinates. Endpoints are accepted only if they lie inside the country geometry. For European countries, the geometry is stabilized to avoid including far-away overseas territories: it is intersected with a Europe-focused bounding box that keeps nearby islands while discarding remote components. For selected states with strong island effects, only the mainland component is retained. This is a modeling decision intended to ensure that the simulated network represents a terrestrial European planning case rather than globally distributed territories that would introduce unrealistic edge lengths and sampling behavior.

\paragraph{Endpoint population model}
Each country has a fixed number of QKD endpoints. Endpoints are interpreted as physical sites where QKD equipment is deployed. Their placement is synthetic and follows two components:
\begin{itemize}
  \item \emph{Cluster component:} for each configured center, typically a capital or major city, a specified number of endpoints is sampled within a circular radius around that center. If $R$ is the cluster radius and $u,v\sim\mathcal{U}(0,1)$, the sampled radial offset is
  \[
  r_{\mathrm{sample}} = R\sqrt{u}, \qquad \theta = 2\pi v,
  \]
  which yields an areally uniform distribution over the disk.
  \item \emph{Rural component:} the remaining endpoints are sampled from a mixture of a uniform draw over the country geometry and a heavy-tailed radial draw around randomly chosen centers. The heavy-tailed radius is
  \[
  r = s\left(u^{-1/\alpha}-1\right),
  \]
  where $s$ is a scale parameter and $\alpha$ controls the tail behavior. This produces many points near existing centers and a smaller number of more distant points, thereby maintaining national coverage.
\end{itemize}
All candidate endpoints are rejected unless they are contained in the country geometry. For multipolygons, the uniform sampler uses area-weighted component selection. Sampling uses bounded attempt limits and fallback strategies to avoid pathological behavior for complex coastlines and fragmented geometries. In Austria, the implementation additionally enforces the dedicated Vienna cluster, the smaller federal-capital clusters, and the St.~Johann im Pongau special cluster.

\paragraph{Distance model}
All distances are computed geodesically on the WGS84 ellipsoid and represent lower-bound proxies for physical fiber length. For local operations such as neighbor ranking, the coordinates are temporarily projected into a metric coordinate system so that Euclidean distance comparisons can be used while preserving relative spatial structure. A detour factor may later be applied when reporting adjusted results in order to approximate the difference between geodesic distance and practical routed fiber distance.

\paragraph{Target degree model and graph semantics}
The network is modeled as an undirected simple graph on the sampled endpoints. Each endpoint is assigned a target degree, that is, a required number of incident edges. The program supports either a manual degree multiset or a constant degree model. In the present study, the degree sequence is specified explicitly per country, typically through counts of degree-2 and degree-3 nodes. Degree targets are treated as hard constraints: a candidate graph is valid only if every endpoint reaches its target degree exactly, with no self-loops and no parallel edges. This encodes a design requirement that each site has a predetermined number of physical fiber connections or feasible QKD links, rather than allowing variable degree from a purely geometric proximity graph.

\paragraph{Edge construction and refinement}
Edge construction is distance-aware but remains constrained by the prescribed degree sequence. The implementation first builds, for each endpoint, a finite set of nearby candidate neighbors in projected space. It then repeatedly connects endpoints with remaining degree demand, preferring short admissible edges and restarting when exact degree satisfaction cannot be achieved. In this sense, the construction is neither a purely geometric nearest-neighbor graph nor a fully unconstrained optimization. It is a repeated search for a simple graph that exactly satisfies the degree sequence while remaining spatially compact.

After a valid degree-feasible graph is found, a degree-preserving local refinement step is applied. Pairs of edges are rewired if the swap preserves simplicity and reduces the total squared projected edge length. At abstraction level, this corresponds to a local optimization of the form
\[
\sum_{(i,j)\in E} \|\mathbf{x}_i-\mathbf{x}_j\|^2 \;\rightarrow\; \min
\]
subject to fixed node degrees and simple-graph constraints. This step does not change the number of edges per node, but it suppresses unnecessarily long links.

\paragraph{Validity constraints beyond degrees}
In addition to exact degree satisfaction, candidate graphs may be required to be connected. Optionally, the algorithm can also reject graphs containing bridges, depending on configuration. These constraints are enforced by rejection and rebuilding: if a candidate violates the required global properties, it is discarded and construction restarts. This matches the planning assumption that the final network must satisfy specific structural properties such as end-to-end reachability between endpoints.

\paragraph{Candidate selection and robustness objective}
For each Monte Carlo run, the program generates multiple valid candidate graphs and selects the one with the best robustness-oriented score. The score is computed from the empirical distribution of edge lengths and is designed to penalize large maximum edge length, high coefficient of variation, and statistical outliers identified through a median-absolute-deviation rule. In abstract form, the implemented score is
\[
S(E)=10^{6}N_{\mathrm{out}}+10^{3}L_{\max}+10^{2}\frac{\sigma_L}{\mu_L},
\]
where $N_{\mathrm{out}}$ is the number of modified-$z$ outlier edges, $L_{\max}$ is the maximum edge length, and $\sigma_L/\mu_L$ is the coefficient of variation of edge lengths. The underlying assumption is that very long or highly uneven link sets are undesirable for cost, deployment feasibility, and operational risk. The selected graph is therefore the valid realization that best suppresses long links and edge-length irregularity within the available candidate set. This is also the criterion used in the generated country reports.

\paragraph{Trusted repeater node model and hop-length interpretation}
After selecting the best endpoint graph, the program places trusted repeater nodes (TRNs) on existing edges under a fixed budget. In our implementation, TRNs are not assigned by a fixed static rule. Instead, the available TRN budget is distributed greedily across edges so as to reduce the largest resulting hop lengths as much as possible. If an edge of length $L_i$ receives $m_i$ trusted repeaters, it is split into $m_i+1$ equal geodesic segments, each of length
\[
h_i = \frac{L_i}{m_i+1}.
\]
At each allocation step, the next TRN is placed on the edge that currently has the largest segment length $L_i/(m_i+1)$, and the repeaters assigned to a given edge are then placed evenly at fractions $j/(m_i+1)$ along that edge, for $j=1,\dots,m_i$. This implements the planning principle that scarce trusted repeaters should be distributed where they most effectively shorten the longest spans. The resulting hop-length statistics after splitting are then reported as the effective QKD span lengths.

\paragraph{Monte Carlo interpretation and reported metrics}
A Monte Carlo run in this framework is not a stochastic failure simulation. It is a stochastic design-space search over feasible degree-constrained graphs induced by random endpoint placements and randomized construction tie-breaking. Each run begins with a new endpoint realization, constructs valid candidate graphs, evaluates them, and retains the best one. The reported distributions across runs therefore reflect variability due to random spatial sampling and graph realization under fixed design rules. Key outputs include total fiber length, hop-length statistics after TRN placement, degree histograms as a verification of constraints, and robustness score components. The implementation also exports per-country edge tables, node tables, plots, configuration files, and summary reports .

\paragraph{Scope and limitations}
The model abstracts away population density fields, existing fiber corridors, terrain constraints, and regulatory siting constraints. Endpoint placement is driven by configured centers and generic rural mixtures rather than measured infrastructure. Graph feasibility is driven by degree constraints and distance preference rather than explicit cost optimization with capacity constraints. Trusted repeater placement is based on edge lengths within the simulated topology rather than detailed facility siting. These choices are consistent with the purpose of synthetic network generation: to produce controlled and comparable instances across countries while enforcing structural constraints and reducing unrealistic long links, especially those induced by remote polygon components.

% Place this after the Simulation section
\begin{tcolorbox}[title=Monte Carlo Approach ,colback=gray!10,colframe=gray!50]

The simulation framework provides a reproducible method for generating national QKD backbone instances.

\begin{itemize}
    \item \textbf{Structure:} endpoints sampled via clustered + rural mixture within country geometry.
    \item \textbf{Topology:} degree-constrained graph with exact node degrees and distance-aware edge selection.
    \item \textbf{Optimization:} multiple candidates evaluated; best graph minimizes long edges and outliers.
    \item \textbf{TRN placement:} greedy splitting of longest edges to reduce maximum hop length.
    \item \textbf{Output:} stable distributions of fiber length and hop statistics from Monte Carlo runs.
\end{itemize}
\end{tcolorbox}

\section{Results for the Austrian Case}
\label{sec:discussion_austria}

This section discusses the results obtained for the Austrian reference case based on the simulation framework described previously. The analysis focuses on key planning metrics, including total fiber length, hop-length characteristics after trusted-node insertion, and robustness indicators. The results are based on $1000$ Monte Carlo runs, where each run produces a valid QKD network realization under the fixed planning assumptions.

\begin{figure*}[ht]
    \centering
    \includegraphics[width=\linewidth]{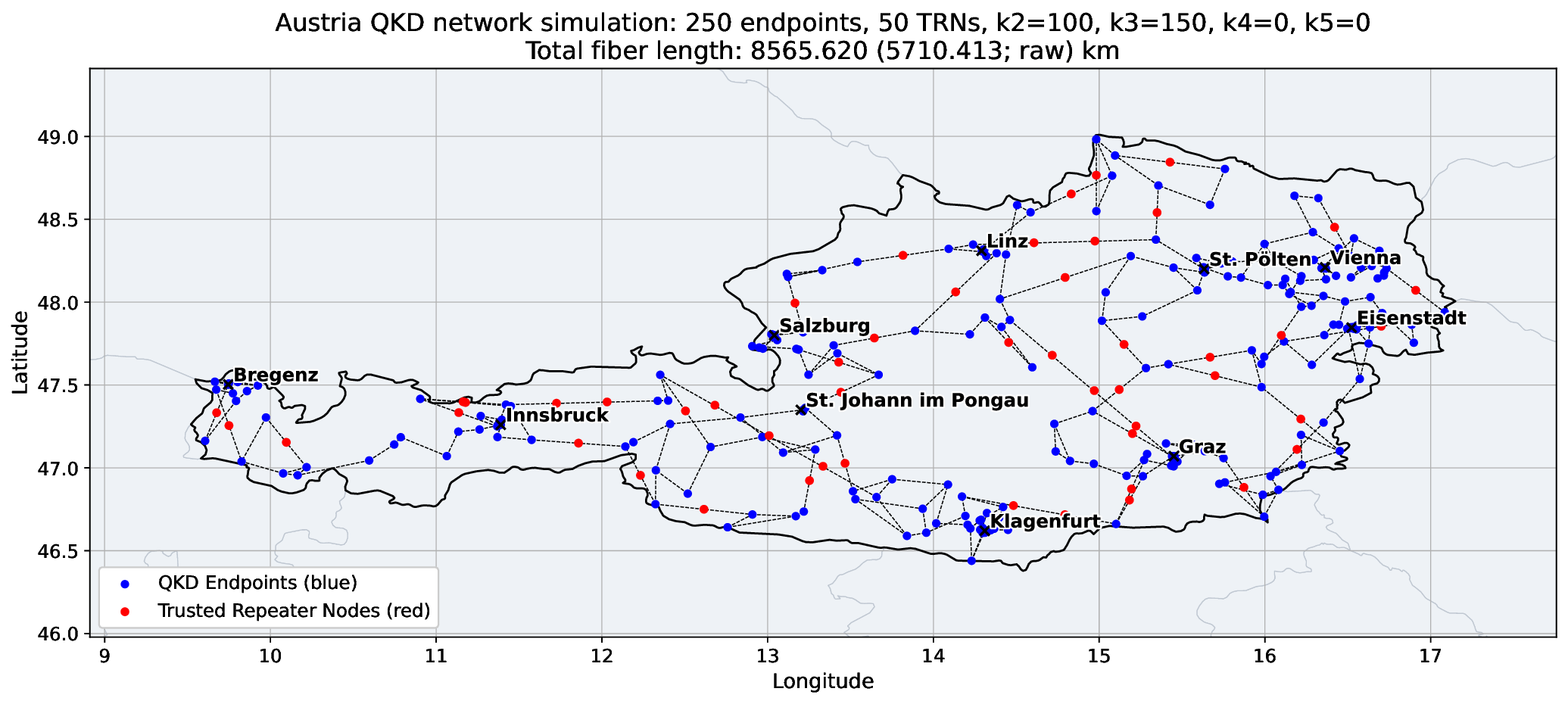}
    \caption{Example realization of the Austrian QKD backbone simulation with $250$ endpoints and $50$ trusted repeater nodes. Blue points represent QKD endpoints, and red points represent trusted repeater nodes.}
    \label{fig:austria_example}
\end{figure*}

Figure~\ref{fig:austria_example} shows one representative simulation instance. Blue markers denote QKD endpoints, while red markers denote trusted repeater nodes (TRNs). The figure illustrates the intended spatial structure of the Austrian case: a stronger concentration in Vienna, smaller clusters around other regional centers, and a distributed rural component across the country. The shown realization has $250$ endpoints, $50$ TRNs, and $325$ QKD links. This confirms that the trusted-node placement strategy effectively reduces longer spans into shorter QKD segments.

Across all Monte Carlo runs, the total fiber length remains stable. The raw value is $5731.03 \pm 7.87$\,km, and the route-adjusted value using $\alpha=1.5$ is $8596.54 \pm 11.80$\,km. The variation across runs is small, which indicates that the total infrastructure size is driven mainly by the national geography and the fixed endpoint count rather than by random placement effects. As a planning result, this suggests that an Austria-wide terrestrial QKD backbone under the chosen assumptions would require on the order of $8600$\,km of routed fiber connectivity.

The hop-length results are favorable with respect to QKD feasibility. After TRN placement, the mean hop length is $15.283 \pm 0.021$\,km raw and $22.924 \pm 0.031$\,km after route adjustment. The maximum hop length is $34.866 \pm 0.056$\,km raw and $52.299 \pm 0.084$\,km adjusted. These values are consistent with the planning assumption that QKD spans should remain in a practically manageable range and, on average, below the internal target of $75$\,km after route correction. At the same time, the results show why trusted nodes are structurally required. Long terrestrial links must be segmented through intermediate trusted sites in order to obtain feasible QKD hop lengths.

The robustness indicators also support the plausibility of the Austrian case. The edge-length coefficient of variation is $0.92 \pm 0.07$, and the number of statistical long-edge outliers is $5.20 \pm 2.53$ per realization. This indicates a moderately heterogeneous but still controlled link-length distribution, which is expected in a country that combines dense metropolitan regions with sparsely distributed alpine and rural areas. The candidate-selection process therefore contributes to suppressing unrealistic long connections while preserving nationwide coverage.

Taken together, the Austrian results provide a consistent planning picture. The total network size is stable, the effective QKD hop lengths remain well within a practical range after trusted-node insertion, and the number of long-edge outliers remains limited. The figure supports this interpretation visually by showing a nationwide but still structured backbone with stronger density in Vienna and regional clusters elsewhere. As a result, the Austrian use case provides a suitable baseline for the later cross-country scaling analysis. In summary, the main quantitative results are: total fiber length of about $5731$\,km raw and $8597$\,km adjusted, mean hop length of about $15.283$\,km raw and $22.924$\,km adjusted, maximum hop length of about $34.866$\,km raw and $52.299$\,km adjusted, and stable behavior across $1000$ Monte Carlo runs.

% Place this after the Austrian results subsection
\begin{tcolorbox}[title=The Austrian Case ,colback=gray!10,colframe=gray!50]
Austria provides a stable planning baseline for national QKD sizing.

\begin{itemize}
    \item \textbf{Scale:} $\sim5731$\,km raw, $\sim8597$\,km adjusted.
    \item \textbf{Nodes:} $250$ endpoints, $50$ trusted repeater nodes.
    \item \textbf{Spans:} mean $\sim15.283$\,km ($22.924$\,km), max $\sim34.866$\,km ($52.299$\,km).
\end{itemize}
\end{tcolorbox}

\section{Expanding the Austrian use case to other EU Member States}
\label{sec:scaling_from_austria_eu}

The Austrian case defined in the previous section is used as the baseline for extending the planning approach to other EU Member States. The main idea is to keep the planning logic unchanged and scale only those quantities that represent national demand and geographic extent. In other words, the assumed QKD architecture remains the same: endpoints represent sites at which secure key material must be available, endpoints are connected through multiple QKD links for basic redundancy, long distances are bridged using trusted repeater nodes, and all participating sites are assumed to host KMS and HSM functionality. What changes from country to country is primarily the number of endpoints and the number of trusted repeater nodes.

The first scaling assumption is that the number of QKD endpoints grows with the size of the protected national demand. At planning level, this demand is approximated through population. This is a simplification, but it is a transparent one. A larger population usually implies a larger number of ministries, agencies, regional authorities, operators of essential services, and other security-relevant facilities that may require access to quantum-generated key material. For this reason, endpoint counts are scaled primarily with population. The second scaling assumption is that the number of trusted repeater nodes is driven more strongly by geographic size than by population. Trusted repeater nodes are introduced to split long terrestrial spans into shorter operational QKD segments. Their number therefore depends mainly on how much national territory must be covered and on how often long inter-regional links arise. For this reason, TRN counts are scaled primarily with land area.

Under this scheme, countries with a large population but limited area tend to require many endpoints but comparatively few trusted repeater nodes. The Netherlands is a clear example. Its endpoint count is increased relative to Austria because the Dutch population is about twice as large, but the TRN count is reduced because the national territory is much smaller. In the current planning configuration, this gives $N_{\mathrm{EP}}=490$ and $N_{\mathrm{TR}}=22$. By contrast, countries with moderate population but very large territory require a different balance. Finland is the clearest case in the present dataset. Its endpoint count is similar to Austria's, but its large territory leads to a strongly increased TRN requirement. In the current planning case, Finland is assigned $N_{\mathrm{EP}}=250$ and $N_{\mathrm{TR}}=100$. This reflects the basic logic of the model: the endpoint layer is driven by demand, while the trusted-node layer is driven by distance.

Large countries such as Germany, France, Italy, Spain, and Poland require increases in both quantities. Germany combines a large population with a large territory and therefore receives a strongly increased planning configuration. France shows a similar pattern, with both a high endpoint count and a high trusted-node count. Spain and France also illustrate an important practical limitation of the terrestrial-only planning approach: geographically complex states with islands or remote maritime territories would require additional assumptions if offshore segments were to be included explicitly. In order to keep the terrestrial comparison consistent, the present simulation framework uses stabilized European country geometries and, for selected countries, restricts the simulation to the mainland component. In the current code, mainland-only handling is explicitly applied to Spain, Portugal, Italy, Greece, Croatia, and France. This means that distant island territories are excluded from the terrestrial planning case rather than being connected through unrealistic maritime trusted-node chains. This simplification is consistent with the scope of this work, which focuses on terrestrial backbone planning and not on offshore or satellite-supported extensions.

The remaining assumptions are kept fixed across all countries in order to preserve comparability. The detour factor remains $\alpha=1.5$ throughout. The trusted-node placement rule also remains unchanged in the sense that trusted repeater nodes are placed only to shorten long terrestrial spans after the endpoint graph has been constructed. In the current implementation, the available trusted repeater budget is distributed greedily across edges so as to reduce the longest resulting hop lengths as effectively as possible. Likewise, the basic redundancy concept is retained: endpoints are still interpreted as sites that should typically support two or three QKD links, and the management layer remains distributed, with one KMS and one HSM per participating site. In this sense, the scaling exercise does not redesign the Austrian logic for each country. It transfers the same planning concept to larger or smaller national settings.

Table~\ref{tab:eu_scaling_from_austria} summarizes the resulting planning inputs for all EU Member States relative to the Austrian baseline. The table should be interpreted as a first-order planning reference, not as a deployment specification. The listed values are suitable as input parameters for the country simulations and for early-stage sizing exercises. They do not replace detailed national analysis, and they do not account for factors such as actual fiber availability, sector-specific criticality, mountains, coastlines, rights-of-way, operator structure, or cross-border integration. Their purpose is narrower and more practical: to provide a consistent and reproducible way to move from one well-defined national reference case to a complete EU-wide set of comparable terrestrial QKD planning scenarios.

\begin{table}[!t]
\centering
\scriptsize
\caption{Scaled planning inputs for extending the Austrian use case to other EU Member States. Endpoint counts are scaled primarily with population, and trusted repeater node counts primarily with land area.}
\label{tab:eu_scaling_from_austria}
\renewcommand{\arraystretch}{1.12}
\setlength{\tabcolsep}{4pt}
\begin{tabular}{lrrrrrr}
\hline
Country & $P$ [M] & $A$ [km$^2$] & $r_P$ & $r_A$ & $N_{\mathrm{EP}}$ & $N_{\mathrm{TR}}$ \\
\hline
Austria      &  9.197 &   83,882 & 1.000 & 1.000 &  250 &  50 \\
Belgium      & 11.900 &   30,667 & 1.294 & 0.366 &  250 &  20 \\
Bulgaria     &  6.437 &  110,996 & 0.700 & 1.323 &  180 &  60 \\
Croatia      &  3.874 &   56,594 & 0.421 & 0.675 &  120 &  30 \\
Cyprus       &  0.980 &    9,253 & 0.107 & 0.110 &   50 &   4 \\
Czechia      & 10.910 &   78,871 & 1.186 & 0.940 &  250 &  50 \\
Denmark      &  5.993 &   42,925 & 0.652 & 0.512 &  200 &  40 \\
Estonia      &  1.370 &   45,336 & 0.149 & 0.540 &   50 &  20 \\
Finland      &  5.636 &  338,411 & 0.613 & 4.034 &  250 & 150 \\
France       & 68.636 &  638,475 & 7.463 & 7.612 &  800 & 200 \\
Germany      & 83.577 &  357,569 & 9.087 & 4.263 & 1000 & 300 \\
Greece       & 10.410 &  131,694 & 1.132 & 1.570 &  280 &  80 \\
Hungary      &  9.540 &   93,012 & 1.037 & 1.109 &  260 &  60 \\
Ireland      &  5.440 &   69,947 & 0.591 & 0.834 &  160 &  40 \\
Italy        & 58.934 &  302,073 & 6.408 & 3.601 &  600 & 150 \\
Latvia       &  1.857 &   64,586 & 0.202 & 0.770 &   50 &  40 \\
Lithuania    &  2.891 &   65,284 & 0.314 & 0.778 &   80 &  40 \\
Luxembourg   &  0.682 &    2,595 & 0.074 & 0.031 &   20 &   4 \\
Malta        &  0.574 &      316 & 0.062 & 0.004 &   20 &   4 \\
Netherlands  & 18.044 &   37,391 & 1.962 & 0.446 &  490 &  22 \\
Poland       & 36.497 &  311,928 & 3.968 & 3.719 &  650 & 200 \\
Portugal     & 10.750 &   92,227 & 1.169 & 1.099 &  260 &  60 \\
Romania      & 19.036 &  238,398 & 2.070 & 2.842 &  550 & 180 \\
Slovakia     &  5.419 &   49,035 & 0.589 & 0.585 &  160 &  30 \\
Slovenia     &  2.131 &   20,273 & 0.232 & 0.242 &   60 &  14 \\
Spain        & 49.078 &  505,983 & 5.336 & 6.032 &  600 & 300 \\
Sweden       & 10.588 &  447,424 & 1.151 & 5.334 &  280 & 180 \\
\hline
\end{tabular}
\end{table}

\subsection{Simulation results for the remaining EU Member States}
\label{sec:eu_results}

The Austrian reference case can now be used to interpret the results for the remaining EU Member States. The same planning logic is retained throughout: endpoints represent sites at which key material must be available, long terrestrial spans are shortened by trusted repeater nodes, and the reported results focus on total backbone length and effective hop lengths after trusted-node placement. What changes from country to country is therefore mainly the overall scale of the network. Countries with large populations tend to require more endpoints, while countries with large territorial extent tend to require more trusted repeater nodes and longer backbone lengths. This produces the expected contrast between, for example, the Netherlands and Finland. The Netherlands has a relatively high endpoint count but short distances and therefore low mean and maximum hop lengths, whereas Finland requires fewer endpoints but many more trusted nodes because of its large territory and long inter-regional distances.

The simulation results reflect this logic clearly. Compact countries such as Luxembourg, Cyprus, Belgium, the Netherlands, and Slovenia show comparatively small backbone lengths and short effective hops. Mid-sized central European states such as Austria, Czechia, Hungary, Slovakia, and Croatia fall into an intermediate range, with route-adjusted mean hop lengths mostly around $20$--$30$\,km and route-adjusted maximum hop lengths typically around $45$--$70$\,km. Larger or geographically more demanding countries such as France, Germany, Italy, Spain, Sweden, Finland, Greece, Poland, and Romania show substantially longer backbone lengths and higher maximum hop lengths, even after trusted-node placement. In these cases, the increase in trusted repeater nodes is necessary to keep the network within a plausible operational range for terrestrial QKD.

For cross-country comparison, Table~\ref{tab:eu_results_single} reports the adjusted and raw result metrics in compact form. In the present result set, the reported route-adjusted maximum hop lengths remain within a practical range of approximately $20$--$95$\,km. This makes the table suitable for comparative planning because it allows country cases to be compared on a common basis without being dominated by isolated geometric extremes. At the same time, the reported values still preserve the structural differences between compact, intermediate, and large national cases. The table therefore provides a useful planning-oriented summary of the effective operational range implied by the current simulation setup.

A second general pattern is that total backbone length increases strongly with national scale, but not in strict proportion to population alone. Germany and France have the largest adjusted backbone lengths in the present comparison, followed by Spain and Italy. Sweden and Finland are also prominent despite smaller populations, because large geographic extent drives longer terrestrial coverage requirements. By contrast, small and compact states remain well below these values. This confirms that endpoint demand and distance demand are distinct drivers and should not be represented by a single scaling variable. It also supports the decision to scale endpoints with population and trusted repeater nodes with land area, while allowing country-specific input configurations where required by geography or national structure.

The results should be interpreted as planning-level estimates and not as deployment blueprints. In particular, the mainland-only simplification remains important for countries with islands or offshore territories. In the present framework, Spain, Portugal, Italy, Greece, Croatia, and France are treated in mainland-focused form for the terrestrial planning case. This avoids unrealistic maritime trusted-node chains and keeps the comparison aligned with the scope of the paper. Within this scope, the overall picture is consistent: the Austrian baseline can be expanded to the rest of Europe in a structured way, and the resulting country cases differ in the expected direction with respect to backbone size, effective hop lengths, and trusted-node requirements.

\begin{table}[H]
\centering
\scriptsize
\setlength{\tabcolsep}{4pt}
\renewcommand{\arraystretch}{1.12}
\caption{Summary of simulation results for EU Member States. For each country, the first line reports the adjusted value followed by the raw value in parentheses, and the second line reports the corresponding Monte Carlo standard deviation as $\pm$ adjusted value followed by $\pm$ raw value in parentheses. Route-adjusted maximum hop lengths are reported as planning values within a practical operational range for terrestrial QKD. Malta is omitted because our simulation did not produce reasonable results due to its small area. 
Here, $\overline{h}$ denotes the mean hop length, where the overline indicates an average over all simulated hops. The subscript $\mathrm{adj}$ refers to values obtained after route adjustment, while $\mathrm{raw}$ refers to values computed directly from the unadjusted simulation. The quantity $h_{\max}$ denotes the maximum hop length observed in the network. The quantity $L_{\mathrm{grid}}$ denotes the total grid length, i.e., the cumulative length of all links in the network. The symbol $\pm$ denotes the Monte Carlo standard deviation, representing statistical variability across simulation runs.}
\label{tab:eu_results_single}
\begin{tabular}{lrrr}
\hline
Country & $\overline{h}_{\mathrm{adj}}$ ($\overline{h}_{\mathrm{raw}}$) & $h_{\max,\mathrm{adj}}$ ($h_{\max,\mathrm{raw}}$) & $L_{\mathrm{grid,adj}}$ ($L_{\mathrm{grid,raw}}$) \\
\hline
Austria     & 22.924 (15.283) & 52.299 (34.866) & 8596.543 (5731.029) \\
            & $\pm 0.995$ ($\pm 0.663$) & $\pm 2.653$ ($\pm 1.769$) & $\pm 373.084$ ($\pm 248.722$) \\
Belgium     & 13.888 (9.259) & 43.955 (29.303) & 4791.515 (3194.343) \\
            & $\pm 0.686$ ($\pm 0.457$) & $\pm 3.091$ ($\pm 2.061$) & $\pm 236.558$ ($\pm 157.705$) \\
Bulgaria    & 27.186 (18.124) & 55.750 (37.166) & 7992.646 (5328.430) \\
            & $\pm 1.461$ ($\pm 0.974$) & $\pm 3.183$ ($\pm 2.122$) & $\pm 429.656$ ($\pm 286.437$) \\
Croatia     & 26.864 (17.909) & 58.473 (38.982) & 4996.667 (3331.112) \\
            & $\pm 1.809$ ($\pm 1.206$) & $\pm 4.736$ ($\pm 3.157$) & $\pm 336.483$ ($\pm 224.322$) \\
Cyprus      & 12.802 (8.534) & 34.485 (22.990) & 883.306 (588.871) \\
            & $\pm 1.402$ ($\pm 0.934$) & $\pm 4.581$ ($\pm 3.054$) & $\pm 96.706$ ($\pm 64.471$) \\
Czechia     & 20.752 (13.835) & 49.800 (33.200) & 7782.015 (5188.010) \\
            & $\pm 0.995$ ($\pm 0.663$) & $\pm 2.620$ ($\pm 1.747$) & $\pm 373.192$ ($\pm 248.795$) \\
Denmark     & 20.572 (13.715) & 51.351 (34.234) & 6171.586 (4114.391) \\
            & $\pm 1.497$ ($\pm 0.998$) & $\pm 4.157$ ($\pm 2.771$) & $\pm 449.164$ ($\pm 299.443$) \\
Estonia     & 29.440 (19.627) & 58.818 (39.212) & 2502.440 (1668.294) \\
            & $\pm 3.479$ ($\pm 2.320$) & $\pm 7.318$ ($\pm 4.879$) & $\pm 295.752$ ($\pm 197.168$) \\
Finland     & 34.607 (23.071) & 62.735 (41.823) & 16438.168 (10958.779) \\
            & $\pm 1.719$ ($\pm 1.146$) & $\pm 3.380$ ($\pm 2.253$) & $\pm 816.513$ ($\pm 544.342$) \\
France      & 27.903 (18.602) & 70.893 (47.262) & 34599.570 (23066.380) \\
            & $\pm 0.828$ ($\pm 0.552$) & $\pm 2.163$ ($\pm 1.442$) & $\pm 1027.145$ ($\pm 684.763$) \\
Germany     & 19.585 (13.057) & 46.673 (31.115) & 31335.630 (20890.420) \\
            & $\pm 0.526$ ($\pm 0.351$) & $\pm 1.348$ ($\pm 0.899$) & $\pm 842.309$ ($\pm 561.540$) \\
Greece      & 25.172 (16.781) & 53.955 (35.970) & 11176.243 (7450.829) \\
            & $\pm 1.121$ ($\pm 0.747$) & $\pm 2.675$ ($\pm 1.783$) & $\pm 497.536$ ($\pm 331.691$) \\
Hungary     & 22.004 (14.670) & 49.657 (33.105) & 8757.735 (5838.490) \\
            & $\pm 0.952$ ($\pm 0.635$) & $\pm 2.387$ ($\pm 1.592$) & $\pm 378.838$ ($\pm 252.559$) \\
Ireland     & 24.313 (16.208) & 54.838 (36.559) & 6029.556 (4019.704) \\
            & $\pm 1.435$ ($\pm 0.957$) & $\pm 3.467$ ($\pm 2.312$) & $\pm 355.975$ ($\pm 237.317$) \\
Italy       & 25.070 (16.713) & 59.075 (39.383) & 23314.851 (15543.234) \\
            & $\pm 0.864$ ($\pm 0.576$) & $\pm 2.211$ ($\pm 1.474$) & $\pm 803.146$ ($\pm 535.430$) \\
Latvia      & 27.218 (18.146) & 42.904 (28.603) & 2857.918 (1905.278) \\
            & $\pm 2.285$ ($\pm 1.523$) & $\pm 3.653$ ($\pm 2.436$) & $\pm 239.943$ ($\pm 159.962$) \\
Lithuania   & 25.480 (16.986) & 45.752 (30.502) & 3669.074 (2446.049) \\
            & $\pm 1.820$ ($\pm 1.213$) & $\pm 3.545$ ($\pm 2.363$) & $\pm 262.038$ ($\pm 174.692$) \\
Luxembourg  & 12.529 (8.353) & 24.272 (16.181) & 375.874 (250.582) \\
            & $\pm 1.708$ ($\pm 1.139$) & $\pm 4.474$ ($\pm 2.983$) & $\pm 51.240$ ($\pm 34.160$) \\
Netherlands & 11.589 (7.726) & 46.072 (30.715) & 7636.855 (5091.237) \\
            & $\pm 0.443$ ($\pm 0.295$) & $\pm 3.263$ ($\pm 2.175$) & $\pm 291.741$ ($\pm 194.494$) \\
Poland      & 24.505 (16.337) & 52.578 (35.052) & 25607.910 (17071.940) \\
            & $\pm 0.682$ ($\pm 0.455$) & $\pm 1.633$ ($\pm 1.089$) & $\pm 713.090$ ($\pm 475.394$) \\
Portugal    & 22.529 (15.019) & 51.112 (34.075) & 8966.465 (5977.644) \\
            & $\pm 0.997$ ($\pm 0.664$) & $\pm 2.599$ ($\pm 1.733$) & $\pm 396.665$ ($\pm 264.443$) \\
Romania     & 22.297 (14.865) & 48.514 (32.343) & 19955.680 (13303.787) \\
            & $\pm 0.701$ ($\pm 0.468$) & $\pm 1.727$ ($\pm 1.152$) & $\pm 627.736$ ($\pm 418.490$) \\
Slovakia    & 20.921 (13.947) & 48.668 (32.445) & 4979.195 (3319.463) \\
            & $\pm 1.183$ ($\pm 0.788$) & $\pm 3.107$ ($\pm 2.071$) & $\pm 281.473$ ($\pm 187.649$) \\
Slovenia    & 19.409 (12.939) & 42.128 (28.085) & 1785.591 (1190.394) \\
            & $\pm 1.618$ ($\pm 1.079$) & $\pm 4.199$ ($\pm 2.800$) & $\pm 148.854$ ($\pm 99.236$) \\
Spain       & 27.783 (18.522) & 54.175 (36.117) & 30005.935 (20003.957) \\
            & $\pm 0.898$ ($\pm 0.599$) & $\pm 1.841$ ($\pm 1.227$) & $\pm 969.823$ ($\pm 646.549$) \\
Sweden      & 37.954 (25.303) & 66.886 (44.590) & 20647.222 (13764.814) \\
            & $\pm 1.822$ ($\pm 1.215$) & $\pm 3.398$ ($\pm 2.265$) & $\pm 991.037$ ($\pm 660.691$) \\
\hline
\end{tabular}
\end{table}

\begin{figure*}[htb]
\centering
\begin{minipage}[t]{0.48\linewidth}
\centering
\includegraphics[width=\linewidth]{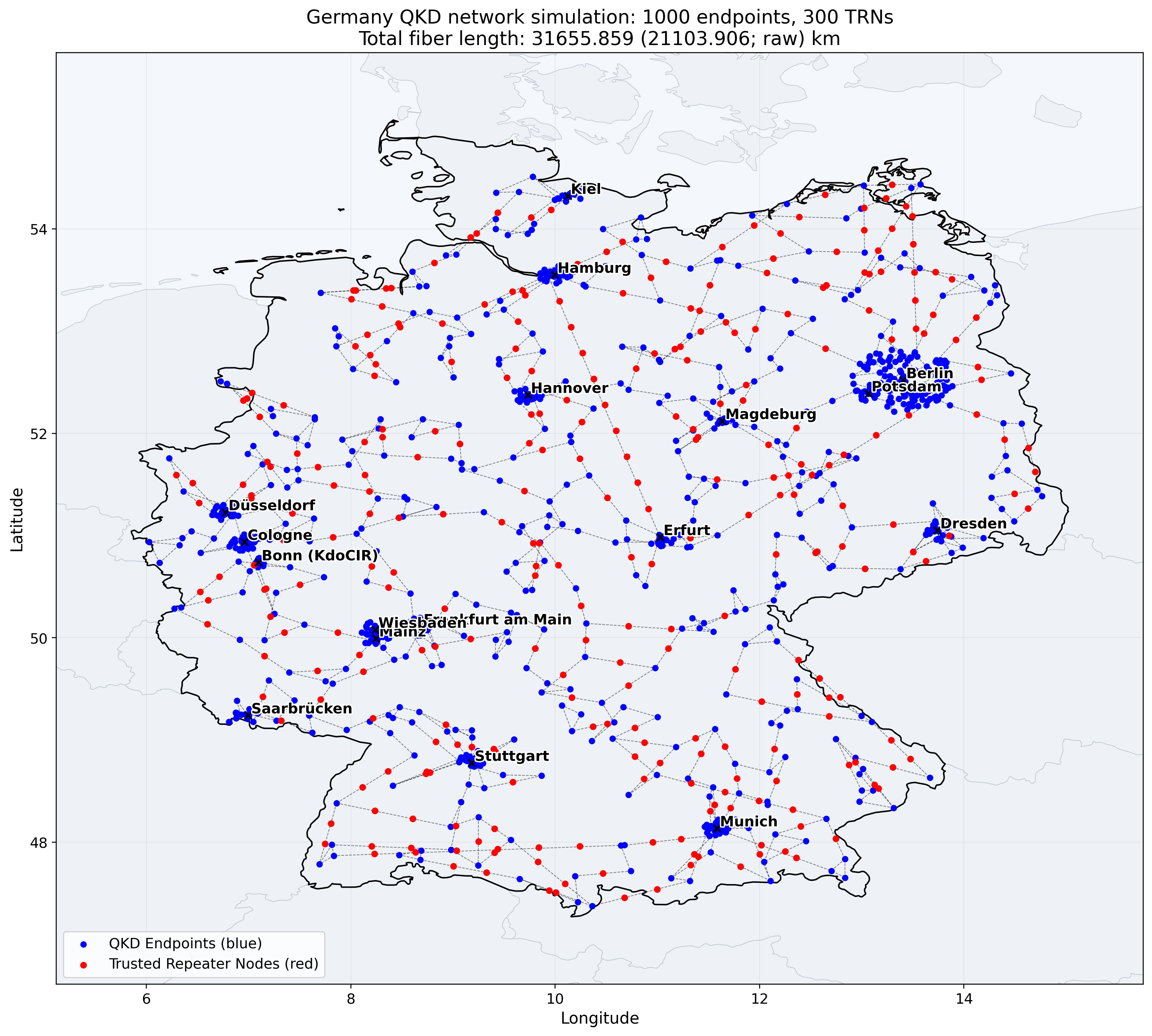}
\caption*{Germany}
\end{minipage}
\hfill
\begin{minipage}[t]{0.48\linewidth}
\centering
\includegraphics[width=\linewidth]{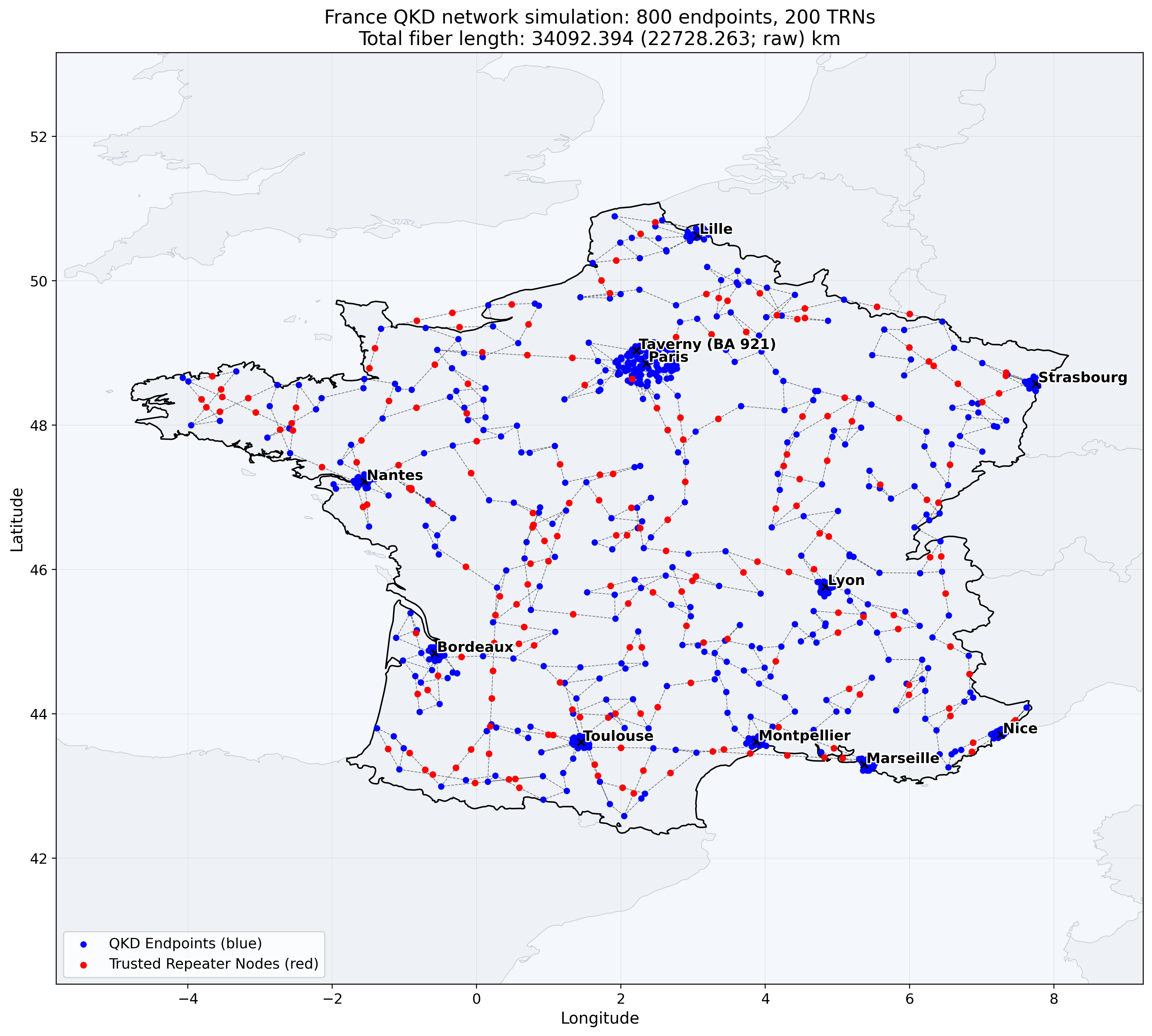}
\caption*{France}
\end{minipage}
\caption{Representative QKD network realizations for Germany and France. Both countries exhibit large-scale backbone structures with wide spatial coverage and a high number of trusted repeater nodes.}
\label{fig:eu_selected_country_plots_1}
\end{figure*}

\begin{figure*}[htb]
\centering
\begin{minipage}[t]{0.48\linewidth}
\centering
\includegraphics[width=\linewidth]{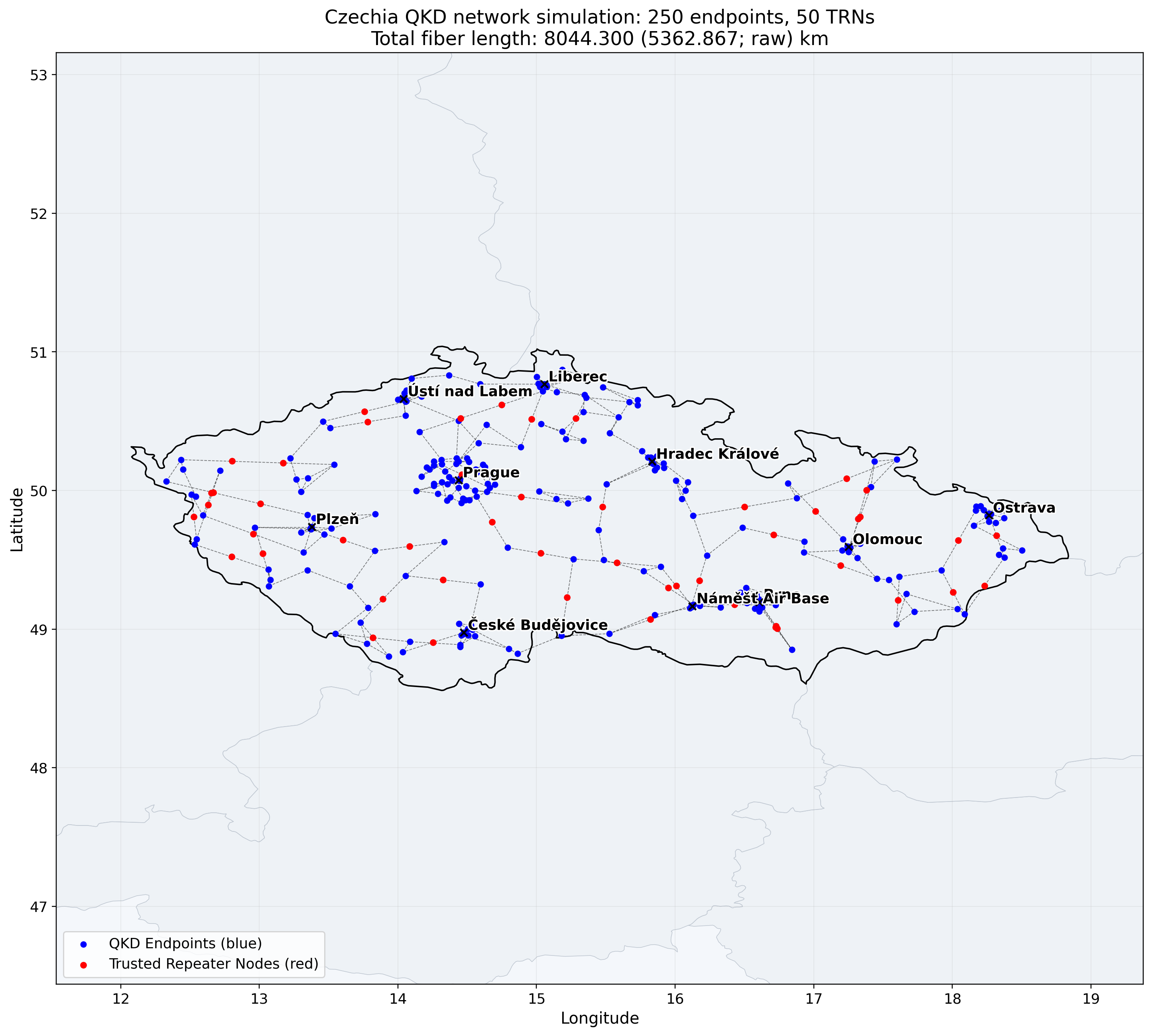}
\caption*{Czechia}
\end{minipage}
\hfill
\begin{minipage}[t]{0.48\linewidth}
\centering
\includegraphics[width=\linewidth]{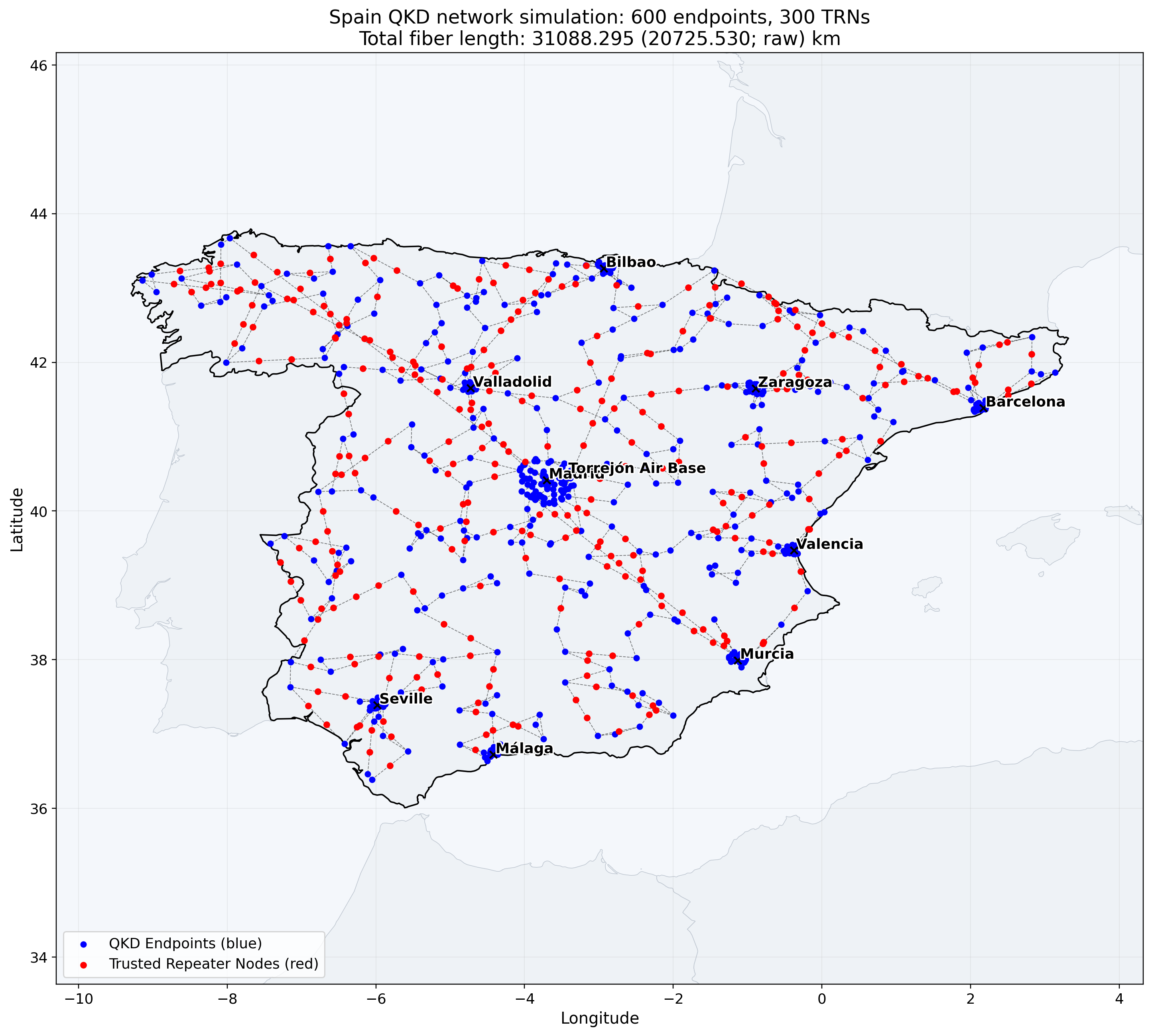}
\caption*{Spain}
\end{minipage}
\caption{Representative QKD network realizations for Czechia and Spain. Czechia shows a compact topology with shorter links, while Spain exhibits a larger geographic spread and longer connections.}
\label{fig:eu_selected_country_plots_2}
\end{figure*}

\begin{figure*}[htb]
\centering
\begin{minipage}[t]{0.48\linewidth}
\centering
\includegraphics[width=\linewidth]{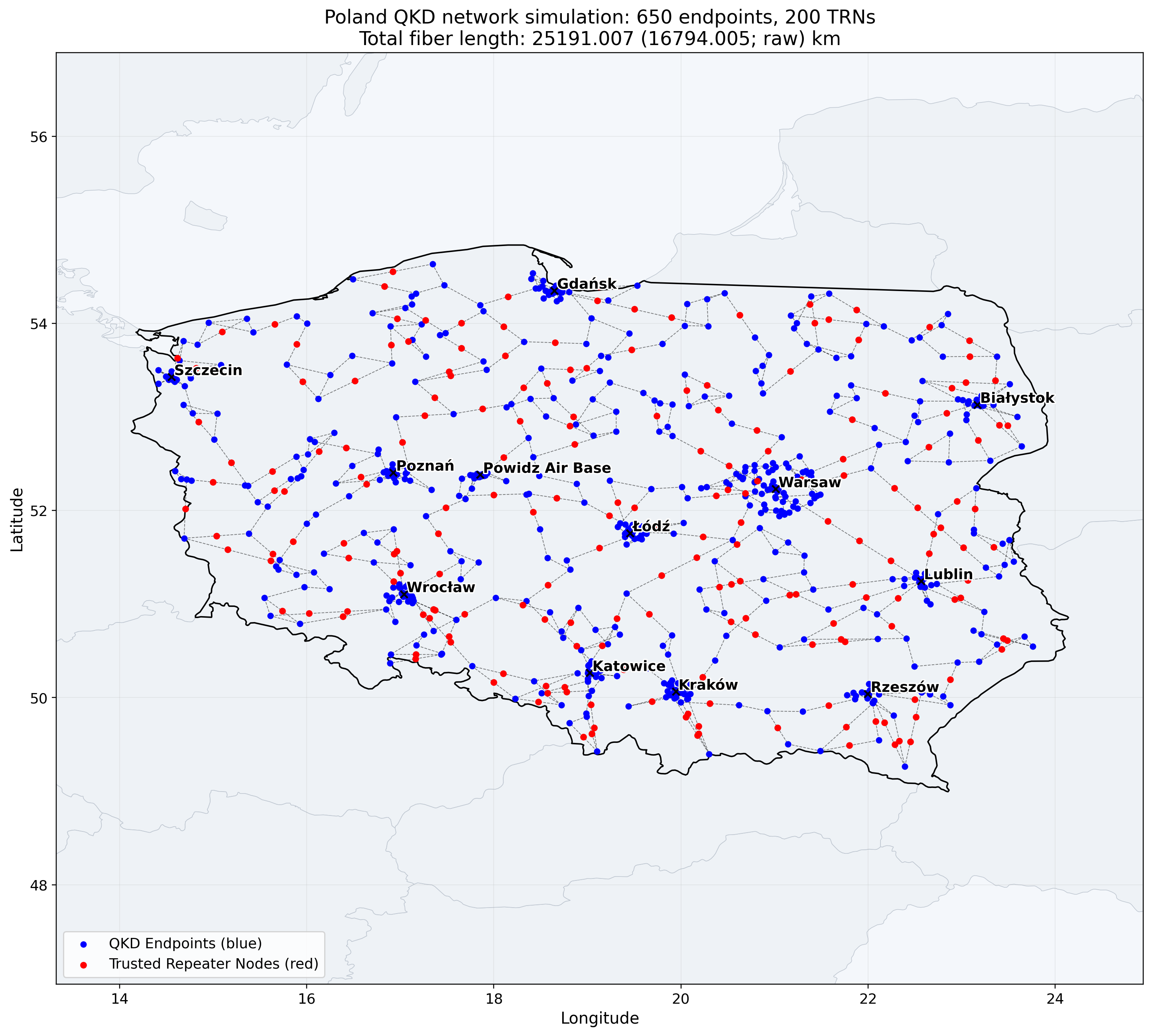}
\caption*{Poland}
\end{minipage}
\hfill
\begin{minipage}[t]{0.48\linewidth}
\centering
\includegraphics[width=\linewidth]{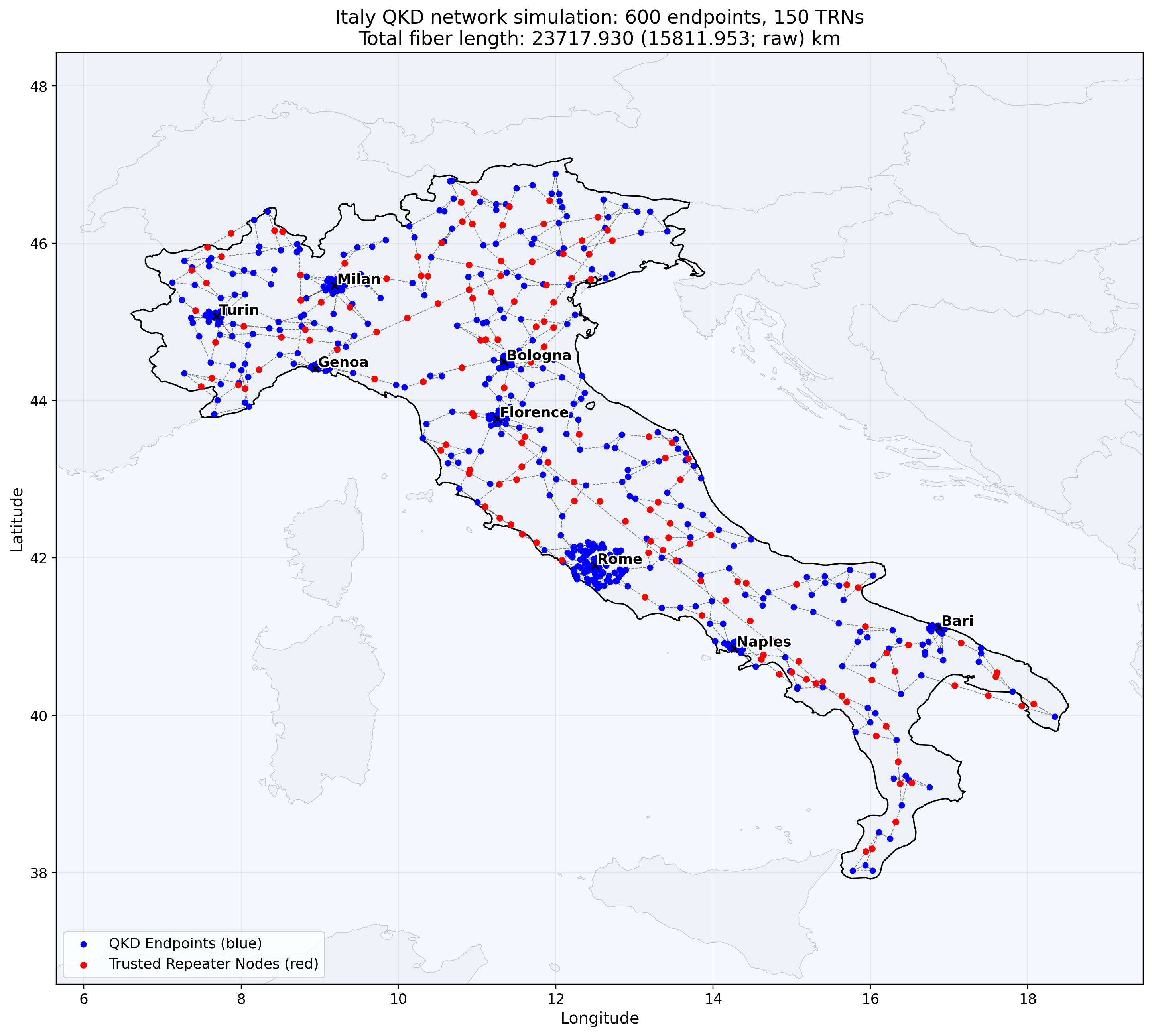}
\caption*{Italy}
\end{minipage}
\caption{Representative QKD network realizations for Poland and Italy. Poland represents an intermediate case, while Italy shows extended backbone structures due to its geographic extent.}
\label{fig:eu_selected_country_plots_3}
\end{figure*}

To highlight the cross-country differences more clearly, Figures~\ref{fig:eu_selected_country_plots_1}--\ref{fig:eu_selected_country_plots_3} summarize representative simulation realizations for six selected countries: Germany, France, Czechia, Spain, Poland, and Italy. These countries were chosen because they represent different combinations of population size, territorial extent, and resulting QKD backbone scale.

Each figure contains two countries for direct side-by-side comparison. The images correspond to valid network realizations with endpoints and trusted repeater nodes placed according to the methodology described earlier. This visualization complements the numerical results by illustrating how spatial distribution and long-distance connections differ across countries.

Czechia shows a comparatively compact and dense structure with shorter links. Germany exhibits a dense backbone with wide national coverage. France, Spain, and Italy display larger geographic spread and longer connections, requiring more trusted repeater nodes. Poland falls between compact and large-country cases. Overall, the examples confirm that the scaling approach produces structurally differentiated network realizations.

% ================= FIGURE 1 =================

% ================= FIGURE 2 =================

% ================= FIGURE 3 =================

\begin{tcolorbox}[title=Scaling to other EU Member States ,colback=gray!10,colframe=gray!50]
The EU results extend the Austrian baseline to consistent national planning scenarios.

\begin{itemize}
    \item \textbf{Scaling logic:} endpoints $\sim$ population, TRNs $\sim$ area.
    \item \textbf{Small states:} short hops, low backbone length.
    \item \textbf{Mid-scale:} $\sim20$--$30$\,km mean, $\sim45$--$70$\,km max hops.
    \item \textbf{Large states:} longer backbones and higher max hops.
    \item \textbf{Outcome:} consistent, comparable EU-wide planning estimates.
\end{itemize}
\end{tcolorbox}

\section{Discussion, limitations, and conclusion}
\label{sec:discussion_conclusion}

The results of this work should be understood as planning-level estimates for terrestrial QKD backbones in Europe. The main contribution is not a deployment design for a specific operator or ministry, but a reproducible sizing method that translates a small set of explicit assumptions into national-scale estimates for endpoints, trusted repeater nodes, effective hop lengths, total backbone length, and associated management components. In this sense, this work addresses a gap between the strategic objectives of EuroQCI and the still limited availability of concrete national planning references. EuroQCI is explicitly conceived as a combined terrestrial and space-based infrastructure, and the first implementation phase of its terrestrial segment has already been supported at EU level. This gives direct practical relevance to terrestrial sizing questions and confirms that terrestrial backbone planning is not a hypothetical exercise but part of the current European implementation path \cite{EC_EuroQCI,CEF_EuroQCI_CallFiche_2024,EC_EuroQCI_2023_2025,EC_ESA_EuroQCI_2025}.

\paragraph{Strategic role of the terrestrial segment}
A central conclusion of this work is that terrestrial QKD remains strategically important in Europe even if space-based QKD becomes available. The space segment is valuable because it can complement terrestrial networks over long distances and in remote areas, and the planned integration of EuroQCI with IRIS$^2$ shows that the European approach is explicitly hybrid rather than purely fiber-based \cite{IRIS2_EC,IRIS2_EUSPA,EC_ESA_EuroQCI_2025}. At the same time, satellite systems introduce a distinct dependency on space assets, space-ground control chains, and ground stations. ENISA's recent space threat landscape report notes that cyber-attacks on satellites can be launched from the ground and that the attack surface extends across development, deployment, operations, and decommissioning \cite{ENISA_Space_Threat_2025}. For this reason, a terrestrial backbone remains important as the continuously controlled, nationally anchored, and physically inspectable part of a sovereign European secure communications architecture. In other words, the space segment is a complement, not a substitute, for a resilient terrestrial core.

\paragraph{Hybrid QKD--PQC architecture}
A further architectural assumption of the present work concerns the integration of QKD backbones with local communication environments, often referred to as the ``last mile''. While QKD is used to establish high-assurance keys between critical infrastructure sites, it is neither practical nor necessary to extend quantum links down to individual end-user devices or office-level systems. Instead, recent work emphasizes hybrid architectures in which QKD-generated keys are terminated at secure sites and subsequently distributed or consumed via quantum secure cryptographic mechanisms, in particular post-quantum cryptography (PQC). In such approaches, QKD provides a high-security backbone for inter-site key exchange, while PQC is used for local distribution, authentication, session establishment, and application-layer protection within sites \cite{ETSI_QKD_014,ETSI_QKD_015,Mehic2024_KMS_Survey}. This hybrid model reduces deployment complexity, avoids the need for dense QKD device placement, and aligns with current recommendations for combining quantum-safe key establishment methods. It also reflects practical system architectures in which KMS instances interface between QKD links and classical cryptographic services, enabling controlled key delivery to applications without requiring direct quantum connectivity at the network edge \cite{James2023_KMS}. Within the scope of this work, this assumption implies that the modeled QKD backbone provides secure inter-site key transport, while the last meters to end systems are covered by PQC-based mechanisms.

\paragraph{Operational relevance of trusted repeaters}
The simulations also confirm the continuing relevance of trusted repeater architectures. With current technology, long-distance terrestrial QKD at national scale still depends on trusted intermediate sites because direct fiber reach is limited and general-purpose quantum repeaters are not yet available for operational deployment \cite{ETSI_QKD_002,ETSI_QS_Whitepaper,Huttner2022}. In the European context, this implies that national QKD backbones are not only about quantum optical links. They also depend on the secure operation of trusted sites, the protection of node hardware, the orchestration of key flows, and the management interfaces between QKD systems and classical control layers. ETSI work on QKD control and orchestration interfaces shows that KMS integration, SDN control, and orchestration are already recognized as part of practical QKD networking rather than optional additions \cite{ETSI_QKD_015,ETSI_QKD_018}. The same applies to HSM-backed key handling and to secure lifecycle management at the application layer. The practical challenge of a national QKD backbone therefore lies not only in laying fiber or installing optical equipment, but also in building a trusted operational environment around the network.

\paragraph{Threshold secret sharing}
A further extension of the proposed trusted-node architecture is the use of threshold secret sharing for secure distributed storage across selected trusted repeater nodes. In such an approach, sensitive data or key material is split into multiple fragments such that reconstruction is possible only if a predefined threshold of fragments is available, while any number of fragments below that threshold reveals no information about the protected content in the ideal model of perfect secret sharing \cite{Shamir1979}. This is attractive in the present context because terrestrial QKD backbones may already rely on trusted nodes as secure intermediate points. If selected trusted nodes additionally host secret-sharing storage components, the infrastructure could combine secure communication with distributed, redundancy-preserving storage. From a security perspective, this reduces single-site exposure because compromise or loss of an individual node does not by itself reveal the stored data as long as the reconstruction threshold is not met. From an operational perspective, threshold-based storage also improves resilience because data remains recoverable even if one or more storage locations become unavailable. Practical distributed-storage designs further show that secret-sharing and dispersal techniques can support confidentiality together with fault tolerance and long-term robustness in distributed environments \cite{Rabin1989,Fujiwara2016,Buchmann2020,DBLP:conf/cloudcom/LorunserHS15}. In this regard, fragmentiX reports a storage architecture in which data is split into multiple fragments and distributed across separate storage locations, with reconstruction possible only from a sufficient subset of fragments; the company also describes appliance features such as stateless operation and hardware-protected credential handling \cite{fragmentiX_Overview_2018,fragmentiX_ApplicationPrimer_2021}. Related application-level demonstrations in QKD infrastructures further indicate that such distributed and protected storage concepts can be integrated with secure communication workflows for sensitive data handling \cite{Zatoukal2021_OpenQKD_Medical}. For the EuroQCI setting, this suggests that trusted repeater nodes could, in addition to their communication role, serve as anchor points for resilient threshold-based storage.

\paragraph{Security dependencies at network level}
Recent work on large-scale QKD networks further shows that trusted-node architectures introduce structural security dependencies that must be addressed at network level. In particular, multi-path approaches demonstrate that distributing key material across multiple independent routes can reduce the impact of compromised nodes and mitigate single points of failure in trusted-node QKD networks \cite{Valbusa2025_MultipathQKD}. This highlights that network topology, routing strategy, and KMS coordination are tightly coupled aspects of security in large terrestrial QKD backbones.

\paragraph{Integration with SDN and KMS infrastructure}
At the same time, recent field demonstrations confirm that large-scale QKD deployments require close integration between QKD systems, software-defined networking, and key management infrastructure. The DISCRETION project shows that QKD-generated keys can be distributed through SDN-controlled environments and integrated with operational systems such as hardened cipher machines under realistic conditions, including military communication scenarios \cite{Bastos2025_DISCRETION_ICMCIS}. Complementary work emphasizes that SDN provides the flexibility required to manage dynamic QKD network configurations, while KMS functionality ensures that generated keys can be securely delivered and consumed across heterogeneous network segments \cite{Brito2025_DISCRETION_QCNC,DBLP:journals/entropy/SimSK23}.

\paragraph{Implications for scalable QKD infrastructures}
These developments reinforce that large terrestrial QKD networks must be analyzed not only in terms of physical connectivity, but also in terms of control-plane integration, key distribution mechanisms, and resilience against node compromise. In particular, they support the view that distributed KMS architectures and multi-path-capable network designs are necessary components of scalable and operational QKD infrastructures. This perspective is consistent with the assumptions used in the present work, where multiple connections per endpoint and distributed key management are treated as baseline design principles for national-scale networks \cite{Valbusa2025_MultipathQKD,Bastos2025_DISCRETION_ICMCIS,Brito2025_DISCRETION_QCNC}.

\paragraph{Interpretation of the model results}
Within this broader context, the results of the Austrian reference case and the EU-wide scaling exercise should be seen as structurally plausible but intentionally simplified. The model captures several important first-order effects. Larger populations increase the demand proxy for endpoint sites. Larger territory increases the need for trusted repeater nodes and longer backbone lengths. Compact states tend to produce shorter mean and maximum hop lengths, while larger or more elongated states require stronger trusted-node support. These are useful planning insights, and they remain visible even under the simplifications used in this work. At the same time, the model does not claim that real national backbones scale only with population and area. Real deployments also depend on state structure, sectoral criticality, existing backbone corridors, operator concentration, military and civil protection requirements, data-centre geography, and the extent to which QKD is applied at central, regional, or sectoral levels.

\paragraph{Main limitations of the study}
Several limitations therefore need to be stated clearly. First, the endpoint distributions are synthetic and are not based on actual protected-site inventories. Second, the backbone lengths are derived from geodesic distances with a uniform detour factor, not from measured fiber routes or leaseable telecom infrastructure. Third, the model does not include explicit traffic demand, key consumption rates, service classes, or capacity planning. Fourth, the security model assumes trusted repeater nodes without quantifying the additional operational risk introduced by each trusted site. Fifth, the country scaling exercise excludes a number of offshore or island cases in mainland-focused simulations, which is a necessary simplification for terrestrial comparability but not a complete national solution for all Member States. Sixth, the method does not yet represent cross-border backbone integration, although EuroQCI is explicitly a pan-European system and not only a set of isolated national networks \cite{EC_EuroQCI,EC_EuroQCI_2023_2025,EC_CEF_2025}. Finally, this work does not provide cost modelling in the full sense of procurement, operations, refresh cycles, staffing, certification, or regulatory compliance. It provides component counts and infrastructure size proxies, but not a complete techno-economic deployment plan.

\paragraph{Scope and practical use of the results}
These limitations do not invalidate the results, but they define the scope of what has and has not been done. What has been done is the construction of a transparent, reproducible, and Europe-wide planning framework for terrestrial QKD backbone sizing. What has not been done is a real network design for any specific Member State. That distinction is important. The present work is useful because it provides consistent first-order estimates where very little comparable planning guidance currently exists. It is not sufficient on its own for procurement or engineering rollout decisions. This work should therefore be read as an infrastructure-dimensioning study that supports early-stage national and EU-level planning, especially in the current period in which EuroQCI terrestrial deployments, orchestration standards, and secure satellite integration are progressing in parallel \cite{EC_EuroQCI,EC_ESA_EuroQCI_2025,IRIS2_EC}.

\subsection{What needs to be done next}
\label{sec:future_work}

The next step should be to move from synthetic demand proxies to country-specific protected-site datasets. For Austria and for other Member States, this means replacing the abstract endpoint layer with classified or semi-classified inventories of ministries, state authorities, military facilities, energy control centres such as power plants and substations, major telecom sites, selected hospitals, and other essential entities. This would make it possible to test whether the current endpoint scaling rules understate or overstate real demand.

\paragraph{Corridor realism}
A second step is corridor realism. The present model uses geodesic distances and a constant detour factor, which is acceptable for first-order planning but insufficient for deployment studies. Follow-up work should incorporate existing long-haul fiber maps, railway and motorway corridors, mountain crossings, urban aggregation structure, and leasing constraints. This is especially important for alpine states, coastal states, and countries with strong regional asymmetry.

\paragraph{Operational modelling of trusted sites}
A third step is operational modelling of trusted repeater nodes, KMS instances, and HSM-backed control. In practice, trusted repeater nodes are not interchangeable abstract points. They are facilities that require secure housing, physical protection, monitoring, maintenance, and certification. The same applies to KMS and HSM infrastructure. Future work should therefore include node trust classes, availability assumptions, lifecycle management, secure update procedures, and failure and incident handling. ETSI control and orchestration work already provides a useful basis for this next phase \cite{ETSI_QKD_015,ETSI_QKD_018}.

\paragraph{Pan-European integration}
A fourth step is explicit pan-European integration. EuroQCI is not only a set of national systems; it is intended to interconnect Member States. Future work should therefore include cross-border terrestrial links, European backbone corridors, and interfaces to the IRIS$^2$-linked space segment where terrestrial coverage alone is insufficient. This is particularly relevant for peripheral regions, islands, and crisis scenarios in which terrestrial paths may be unavailable or degraded \cite{EC_EuroQCI,IRIS2_EC,IRIS2_EUSPA}. At the same time, such work should continue to treat the space segment as a complement that improves reach and flexibility, not as a reason to weaken terrestrial backbone ambitions.

\paragraph{Techno-economic analysis}
A fifth step is techno-economic analysis. The present article estimates counts and lengths, but the next planning phase should convert these into procurement categories, operational expenditure models, staffing needs, certification effort, and staged deployment plans. That is the point at which national planning can move from ``How large might the backbone be?'' to ``How should it actually be built and operated?''

\begin{resultsummarybox}
The main outcomes of this work provide a structured and consistent basis for planning-level sizing of national terrestrial QKD backbones in Europe.

\begin{itemize}
    \item EuroQCI creates an immediate need for planning-level estimates of national terrestrial QKD backbones, and such estimates are still largely missing in the literature.
    \item A terrestrial QKD backbone remains strategically important in Europe even if space-based QKD is added later, because terrestrial infrastructure offers a more continuously controlled and physically anchored security layer.
    \item National QKD scaling is driven by at least two different factors: endpoint demand is linked to protected-site density, while trusted repeater demand is linked to geographic extent.
    \item Trusted repeater nodes, KMS instances, and HSM-backed key handling are not secondary details; they are central components of any realistic national QKD architecture.
    \item The presented results are useful as first-order planning references for Europe, but they are not deployment designs and must be refined with real site inventories, route data, cross-border integration, and operational cost modelling.
\end{itemize}

\end{resultsummarybox}

\section*{Acknowledgements}
This work was supported by the Digital Europe Programme under project number 101091642 (“QCI-CAT”).

This work was also supported by Open Access Funding provided by the University of Vienna and SBA Research. SBA Research (SBA-K1 NGC) is a COMET Center within the COMET—Competence Centers for Excellent Technologies Programme and is funded by BMIMI, BMWET, and the federal state of Vienna. The COMET Programme is managed by the Austrian Research Promotion Agency (FFG).

The financial support by the Austrian Federal Ministry of Economy, Energy and Tourism, the National Foundation for Research, Technology and Development and the Christian Doppler Research Association is gratefully acknowledged.

\bibliographystyle{alpha}
\bibliography{references}

\end{document}